\documentstyle[osa,12pt]{revtex}

\begin{document}
\title{Surface Action for a Point Particle }
\author{Marius I. Piso, Nicholas Ionescu-Pallas}
\address{Institute of Space Sciences - Gravitational Researches Laboratory\\
21-25 Mendeleev str., 70168 Bucharest, Romania}
\maketitle
\date{}

\begin{abstract}
A remark on the movement of a point mass particle is given. If one
associates to the particle a sphere of radius equal to the related De
Broglie length, the relativistic action on the trajectory is proportional to
the surface described by this sphere.
\end{abstract}


\section{Introduction}

In standard relativistic mechanics, the base space-time is considered to be
the Minkowski flat space, i.e. a real 4-dimensional manifold endowed with
the Lorentzian metric $\eta _{ik}$ \cite{LL2}. A common procedure to derive
the general properties of the movement is the use of the principle of least
action. For a point particle of rest mass $m_0$, the action $S$ may be
written \cite{LL2}: 
\begin{equation}  \label{1}
S=-m_0c\int_i^f(\eta _{ik}dx^idx^k)^{1/2}\equiv -m_0c\int_i^fds\;\sim
\;L_{if}
\end{equation}
as being proportional to the {\em length} $L_{if}$ of the world line of the
point particle in the Minkowski space between some initial $i$ and final $f$
points; $ds$ is the element of length of the world line, $c$ is the speed of
light and $v$ is the speed of the particle. In string theories, the particle
becomes a 1-dimensional object. The simplest type of action is the
Nambu-Goto one \cite{GSW}: 
\begin{equation}  \label{2}
S=T\int_i^f\left[ \dot X^2X^{\prime 2}-(\dot X\cdot X^{\prime })^2\right]
^{1/2}d\sigma d\tau \equiv T\int_i^fd\Sigma \;\sim \;\Sigma _{if}
\end{equation}
which is proportional to the {\em area} $\Sigma _{if}$ of the world sheet of
the string embedded in the Minkowski space; $d\Sigma $ is the element of
area of the world sheet and $T$ is a constant of proportionality introduced
in order make the expression of the dimensions of action. No more details
are needed for the further understanding of the paper. If necessary, see 
\cite{GSW}.

The conclusions of this part is that the dynamic properties of the movement
of the particles may be described by means of an action which is
proportional to geometrical objects attached to the Minkowski space-time
description (the {\em length} for the point particle and the {\em area} for
the string). These concepts does not change in fact when passing from
classical to quantum mechanics.

\section{Surface proportional action for a point particle}

We shall study the movement of a point particle of mass $m$ . In the wave
quantum mechanics, one may associate to this particle the De Broglie
characteristic length $\Lambda _B$ \cite{Fock}: 
\begin{equation}  \label{3}
\Lambda _B=\frac h{2\pi p}=\frac h{2\pi mv}
\end{equation}
where $h$ is the Planck action constant, $p$ - the momentum, $m$ - the mass
and $v$ the velocity of the particle in some inertial frame (let's say the
laboratory frame).

From the quantum mechanical point of view, we may consider in a good
approximation that the particle is confined in a spatial zone delimited by a
sphere whose radius is just the De Broglie length $\Lambda _{B\text{. }}$In
this approximation, we are allowed to introduce a sphere of radius $\Lambda
_B$ attached to the particle. If the particle moves onto the direction $x$,
this sphere determines an axis symmetric surface around the $x$ axis in the
laboratory frame. The area of this surface between the positions $x_i$ and $%
x_f$ is, obviously: 
\begin{equation}  \label{4}
A_{if}=2\pi \int_{x_f}^{x_f}\Lambda _B\cdot dx=2\pi \int_{x_i}^{x_f}\frac h{%
2\pi mv}\cdot dx=\int_{x_i}^{x_f}\frac h{mv}\cdot dx
\end{equation}
By replacing $dx=v\cdot dt$ and taking into account the relativistic mass
dependence on speed \cite{LL2}, $A_B$ becomes: 
\begin{equation}  \label{5}
A_{if}=\int_{t_i}^{t_f}\frac hm\cdot dt=\frac h{m_0}\int_{t_i}^{t_f}\left( 1-%
\frac{v^2}{c^2}\right) ^{1/2}\cdot dt
\end{equation}
where $m_0$ is the rest mass of the particle and $c$ - the speed of light.

On the other side, the relativistic action \ref{1} for the same point mass
particle is: 
\begin{equation}  \label{6}
S_{if}=-m_0c\int_i^fds=-m_0c^2\int_i^f\frac{ds}c=-m_0c^2\int_{t_i}^{t_f}%
\left( 1-\frac{v^2}{c^2}\right) ^{1/2}\cdot dt
\end{equation}
From \ref{5} and \ref{6}, we are able to assume that the relativistic action
for a particle of rest mass $m_0$ is: 
\begin{equation}  \label{7}
S_{if}=\frac{m_0^2c^2}h\cdot A_{if}=\frac h{\lambda ^2}\cdot A_{if}
\end{equation}
proportional to the area $A_{if}$ of the spatial surface generated by the De
Broglie length radius sphere; $\lambda $ is the equivalent Compton length
for the particle of rest mass $m_0$.

\section{Conclusions}

There are two general distinct methods to describe a particle: point
particle or string. The dynamics of a point particle is described by a world
line length proportional action, meanwhile the string action is proportional
to the area of the world sheet. Starting from basic quantum mechanical
assumptions, we considered that a point particle is confined in the interior
of a sphere of radius equal to the attached De Broglie wavelength. After
simple calculations, we proved that the area of the spatial surface
described by this sphere attached to the particle is proportional to the
standard relativistic action. This toy model could establish an intuitive
argument for the progress of peace in the particles vs. strings war.

\end{document}